\documentclass[a4paper,11pt]{article}
\usepackage{pos}
\usepackage{url}
\title{Status and perspectives of ILDG}
\author*[a]{Christian Schmidt}

\affiliation[a]{Fakult\"at f\"ur Physik, Universit\"at Bielefeld,\\
  Universit\"atsstrasse 25, 33615 Bielefeld, Germany}

\emailAdd{schmidt@physik.uni-bielefeld.de}

\abstract{We discuss the status and progress of recent efforts to modernize the International Lattice Data Grid (ILDG). This includes activities of the metadata and middleware working groups concerning deployment and operation of crucial services (user management, metadata catalogues, file catalogues) and extensions of the metadata format, which have been tailored according to the needs of the large collaborations. We also report on developments and extensions that are planned to be addressed in the foreseeable future.}

\FullConference{The 42nd International Symposium on Lattice Field Theory (LATTICE2025)\\
2-8 November 2025\\
Tata Institute of Fundamental Research, Mumbai, India\\}

\begin{document}
\maketitle

\section{Introduction}
The International Lattice Data Grid (ILDG) \cite{Davies:2002mu, Irving:2003uk, Ukawa:2004he, Beckett:2009cb, ildg-organization} is an initiative with the purpose of enabling and coordinating the sharing of research data within the research community in Lattice Field Theories.
The ILDG is organized as a Virtual Organization (VO). It provides a central user management
service \cite{IAM} and defines the specifications of the metadata schemata, the file format, and the interface of the catalog services. In turn, the actual storage resources and catalog services are provided and managed in an autonomous way by the various regional grids. 
Traditionally, the grids have been CSSM \cite{CSSM} for Australia, JLDG \cite{JLDG} for Japan, Latfor Data Grid (LDG) \cite{LDG} for continental
Europe, UK Lattice Field Theory for the UK \cite{UKLFT}, and USQCD \cite{USQCD} for the US. The ILDG has two working groups, the metadata working group (MDWG), which specifies and develops the metadata standard and file format, and the middleware working group (MWWG), which defines and coordinates the implementation of services and middleware for the interoperability of the regional grids. These activities are overseen by the ILDG board. 
Since 2021, the ILDG benefits from the German research data infrastructure initiative (NFDI). Within one of its consortia the Particles, Universe, NuClei and Hadrons for the NFDI (PUNCH4NFDI), the java code for the catalog services has been refactored.

%%%
%%%

\section{Recent highlights}
The International Lattice Data Grid (ILDG) has recently completed a major upgrade and modernization, referred to as ILDG~2.0, which is now fully operational and ready for use.
Key components, like the federated user management service and the metadata catalogs have been replaced or modernized, and the metadata schema has been extended. The important changes are discussed below. 

\subsection{The Identity and Access Management (IAM)}
Users and VO membership are now managed through the INDIGO Identity and Access Management (IAM)
service, which is developed and maintained by INFN-CNAF. The dedicated IAM instance for ILDG
is also hosted at CNAF and currently there are $\mathcal{O}(100)$ ILDG members registered in the IAM.  
It is configured with Single-Sign-On, using the eduGAIN federation of identity providers (IdP). 
This means that all users at login to the IAM are redirected to the IdP of their home institutions, where they can authenticate with their local credentials. This completely eliminates the need for
users to hold a valid Grid certificate, which in some cases was a major obstacle for using the ILDG.
Also authentication and authorization towards the services is now fully based on tokens.
With this significantly simplified user registration and a fine-grained token-based access control,
two major milestones in the ILDG agenda have been reached.

\begin{figure}
    \centering
    \includegraphics{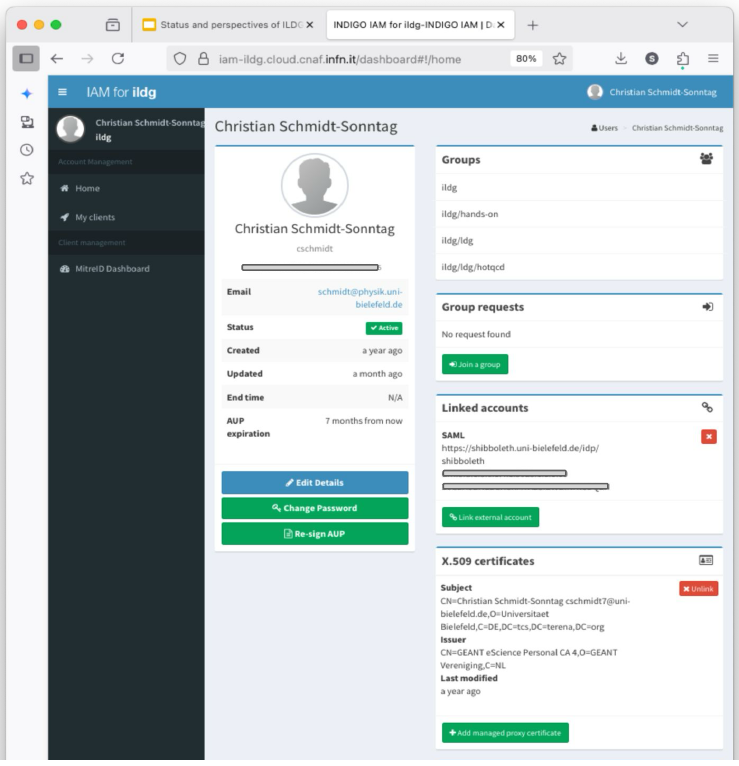}
    \caption{Dashboard of an ILDG user registered in the IAM}
    \label{fig:IAM}.
\end{figure}

When requesting ILDG membership at the IAM, users are required to accept the VO policy \cite{vo_policy}
and an Acceptable Use Policy (AUP) \cite{AUP}.
To retain the membership, the acceptance has to be repeated periodically every twelve months. 
After login to the IAM, the user will see a dashboard, as shown in Fig.~\ref{fig:IAM}.
Besides information on the user's identity and the button to re-sign the AUP, the user is able to
create clients that can receive capability tokens from the IAM. In turn, these tokens provide
authorization to access storage elements for reading or writing of configurations, or to register metadata in the catalogs. 
The scope of the tokens defines the detailed read- and write permissions of the user for specific resources, e.g. in certain directories. The allowed scopes can be derived from the group memberships of the user, which are also managed in the IAM and visible in the dashboard. 
Finally, we note the possibility to link the account of the user to an X.509 grid certificate,
which can then be used as an alternative and convenient authentication method for the IAM login. 

\subsection{Catalog services}
\begin{figure}
    \centering
    \includegraphics[width=0.8\textwidth]{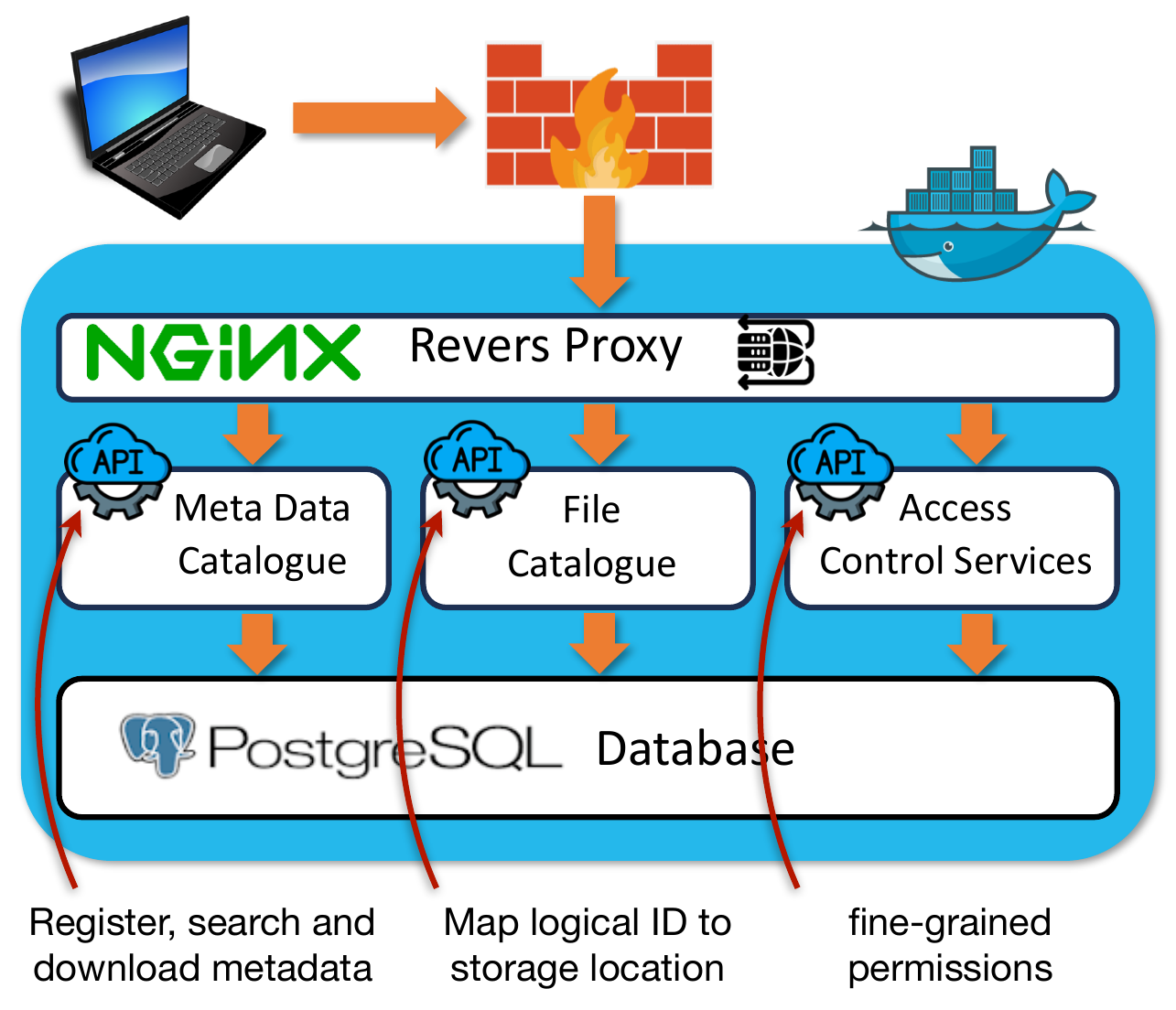}
    \caption{Structure of a containerized implementation of ILDG web services.}
    \label{fig:services}
\end{figure}

ILDG uses two kinds of catalogs: metadata catalogs contain all metadata
that describes the ensembles and configurations, while file catalogs map
the logical ID of configurations to their (possibly replicated) storage locations.
The interface specifications of the metadata and file catalogs in ILDG 2.0 have
been updated to a modern (REST) API. They include complex
search operations and downloading of metadata, as well as optional extensions
for write-access with fine-grained token-based authorization. A corresponding
reference implementation of the catalogs, which can also be deployed as
stand-alone services, is available as Docker containers. Their general structure
is shown in Fig.~\ref{fig:services}.
% Further technical aspects include the transition to PostgreSQL as database backend to
% improve search performance, and the containerization to allow simple deployment at mutliple
% sites or regional grids.
Containerization and making use of the same implementation of the catalogs, as currently
done by the LDG, JLDG and UKQCD regional grids, considerably simplified the deployment
and operation of these services, and it may help to join efforts for their long-term maintenance.

An additional access control service can be used to define very fine-grained authorization
policies in a hierarchical way. Currently, this feature is not yet used and authorization
decisions are directly managed by scope policies defined within the IAM. The coherent
and fully token-based access control to all (meta-)data resources is an important new
feature in ILDG 2.0, because it enables collaborations to keep uploaded data under
embargo for some limited amount of time, i.e. to reliably restrict access to the
configurations to internal use only, before they are made public.

We also note that the new implementation of the metadata catalog is completely agnostic to
the used metadata schemata. This makes it an attractive solution also for other communities.
As part of the work program in the second funding period of PUNCH4NFDI it is planned to realize ILDG-like data management frameworks also for other scientific communities. 

\subsection{Metadata schema and file format}
A revised and extended version 2.0 of the QCDml metadata schemata for ensembles and configurations has been released.
It incorporates the requirements from several lattice QCD collaborations and enables them to markup their configurations according to their needs. Most importantly, the schema is now fully FAIR compliant \cite{FAIR}, where FAIR stands for "Findable, Accessible, Interoperable and Reusable". It now includes a mandatory license statement, optional funding information, and the possibility to indicate embargo periods.
Several new lattice actions, e.g. QCD+QED, gauge groups beyond SU(3), and open boundary conditions are now also supported. 
In addition, the format specifications for data files have been extended to support a compressed data format and packing of multiple configurations into a single LIME file.

\section{Interacting with ILDG services}
Simple command line tools to interact with the catalogs for the purpose of uploading, searching, and downloading of metadata are available as repository or within a container \cite{hands-on}. They are basically wrapper scripts using \textit{curl} \cite{curl} commands to generate suitable
HTTP requests and do not require installation of any complex grid software.
More high-level tools to perform the entire upload workflow with all required write
operations to metadata catalog, storage elements, and file catalog are still desirable
(possibly also for download operations, although these are simpler).
With the API of the catalogs being documented (and accessible) through the Swagger interface
\cite{swagger}, it is also straight forward for users to develop their own customized tools,
web pages, or scripts for the interaction with ILDG services. Several examples of web interfaces are now online,
ranging from simple listings \cite{simple} to the reactivated faceted navigation page of JLDG
and the newly developed markup and search interfaces \cite{markupGUI,searchGUI}.
An add-on service for metadata harvesting via OAI-PMH (Open Archives Initiative Protocol
for Metadata Harvesting) is also available to integrate ILDG data into other research
data services, like INSPIRE or the PUNCH science data platform. 

Storage elements are accessed through standard protocols (typically also HTTP or WebDAV)
with a broad choice of existing client tools. ILDG currently has a total
of six storage elements in the various regional grids: four in LDG (DESY Hamburg and Zeuthen, J\"ulich Supercomputing Centre, and INFN-CNAF), one in Japan (Tsukuba University) and one in the UK (Swansea University). They are configured to support at least read access with tokens issued by the IAM of ILDG.
In total about 400 ensembles with ~350,000 configurations (mostly legacy data from
ILDG 1.0) are currently stored on the ILDG storage elements. Various collaborations
have plans for massive uploads of new ensembles to ILDG, see Ref.~\cite{Bali:2022mlg,Aoki:2025etz}.

\section{Outlook}
ILDG 2.0 with its revised and modernized services, infrastructure, and metadata schema
is now ready to be used for new data. Various collaborations have plans for massive
uploads of new ensembles to ILDG, see Ref.~\cite{Bali:2022mlg,Aoki:2025etz}. Starting
such uploads requires some time and preparatory work by the data providers. However, this
effort can be expected to quickly pay off through an improved (even only collaboration
internally) and fully FAIR-compliant data management for the gauge configurations.

Further development efforts by the working groups and user community are expected to
focus on more convenient and high-level client tools with graphical as well as
command-line interfaces. Moreover, a tight monitoring of the status of all services
(IAM, catalogs and storage) is desirable for a highly reliable production mode.
Development plans of the metadata working group include the support for additional file formats, such as HDF5, and the setup of standardized workflows for data publishing, including DOI minting, landing pages, and metadata harvesting.
Community training and on-boarding activities, like the recent hands-on
workshop \cite{hands-on-indico}, need to be repeated for both data providers
and as data consumers.

Looking forward, joint efforts are needed to ensure long-term sustainability of
ILDG activities through adequate funding and manpower commitments, and the lattice
community might want to reassess the structure of regional grids.
In Europe, sustainability questions are currently further addressed and organized through EuroLFT \cite{euroLFT}. In addition, ILDG aims to broaden the impact of its technology beyond lattice QCD towards other scientific communities. These activities are pursued, for instance, within the PUNCH4NFDI consortium and might generate synergies on efficient and FAIR compliant research data management. Continued participation in ILDG working groups is encouraged to plan and realize future developments. For questions or suggestions please contact the ILDG working groups or board through {\tt ildg-contact}(at){\tt desy.de}.

\section*{Acknowledgments}
The status and progress of ILDG 2.0 as reported here has only been possible through
the great effort of members of the ILDG working groups and the support by the ILDG
board. In particular, we point out recent essential contributions and work by
Antonio Rago,
Basavaraja Bheemalingappa Sagar,
Craig McNeile, 
Daniel Kn\"uttel,
Dirk Pleiter,
Ed Bennet, 
Georg von Hippel,
Giannis Koutsou,
Giovanni Pederiva,
Hideo Matsufuru,
Hiroshi Ohno,
Hubert Simma,
Osamu Tatebe,
and
Yoshinobu Kuramashi.
A crucial element of ILDG 2.0 is the INDIGO~\,IAM, and we a grateful to 
the INFN-CNAF team for their great development, support, and help, in
particular to Federica Agostini, Francesco Giaccomini, Roberta Miccoli,
Diego Michelotto, Carmelo Pellegrino, and Enrico Vianello.
We thank the IT teams of all ILDG sites for their support and
help, in particular, at CNAF, DESY, J\"ulich Supercomputing Centre, and
Tsukuba University.
We are acknowledge important support from various members of the PUNCH4NFDI
consortium, in particular, Gunnar Bali, Olaf Kaczmarek, Frithjof Karsch,
Stefan Krieg, Carsten Urbach, and Tilo Wettig. 

I am grateful to Hubert Simma for valuable comments and suggestions to this contribution.
The work of CS is supported by Deutsche Forschungsgemeinschaft (DFG, German Research Foundation) - Project No. 315477589 (CRC-TR 211) and 460248186 (PUNCH4NFDI) as well by the competence network HPC.nrw of the state of North Rhine-Westphalia.
%\begin{thebibliography}{99}
%\bibitem{...}
\bibliographystyle{JHEP}
\bibliography{bib}

%\end{thebibliography}

\end{document}